\documentclass[conference]{IEEEtran}
\IEEEoverridecommandlockouts
\usepackage{cite}
\usepackage{amsmath,amssymb,amsfonts}
\usepackage{algorithmic}
\usepackage{graphicx}
\usepackage{textcomp}
\usepackage{xcolor}
\def\BibTeX{{\rm B\kern-.05em{\sc i\kern-.025em b}\kern-.08em
    T\kern-.1667em\lower.7ex\hbox{E}\kern-.125emX}}
\begin{document}

\title{CloudCap (C$^{2}$app) : A Cloud-Based Platform for Packet Analysis  On The Edge}
% EdgeCap, An Adaptable Mobile Network Capturing Tool On The Edge

\author{Kyriazis Kokkinos,
        Ioannis Polymenidis,
        Ilias Siniosoglou,\\
        Athanasios Liatifis,
        Panagiotis Sarigiannidis
      
\thanks{ K. Kokkinos, I. Polymenidis, I. Siniosoglou, A. Liatifis and P. Sarigiannidis are with the Department of Electrical and Computer Engineering, University of Western Macedonia, Kozani 50100, Greece - \texttt{E-Mail: \{ece01016, ece01058, isiniosoglou, aliatifis, psarigiannidis\}@uowm.gr}
}}
%P Radoglou-Grammatikis pradoglou,
        % Panagiotis Radoglou Grammatikis,
\maketitle

\begin{abstract}
Data exchange through mobile devices is rapidly increasing due to the high information demands of today's applications. The need for monitoring the exchanged traffic becomes important in order to control and optimize the device and network performance and security. Taking this under consideration, in this paper, we developed a cloud-based system for the analysis of network traffic. The smartphone devices act both as traffic captors and visualization end-points, enabling the user to get an overview of the network while minimizing resource consumption. In the presented work, we evaluate our system using two test cases and a variety of target devices. Our results prove the usefulness of the proposed system architecture.

% This application provides the user with the ability to capture the network traffic from the available interfaces of the device and send them to a server for packet analysis. Afterwards, the user can either view the data in a graphical way such as graphs or as a simple list that contains some basic information about each packet.
\end{abstract}

% ========================================================================================================================
% =====================================================Introduction=======================================================
% ========================================================================================================================

\section{Introduction}
\IEEEPARstart{I}{}n the modern technological age, computer networks constitute the basic pillar of machine inter connectivity, data dissemination and interpersonal information exchange and socializing. As technology advances and the contextual requirements of the variety of existing applications and services becomes ever more demanding, computer networks are bombarded with an endless supply of communication bursts between devices. With the introduction of Wireless networks and the advancement of the mobile device paradigm the exchanged traffic in both Wide Area Networks (WANs) and Local Area Networks (LANs) has increased exponentially \cite{MamadouMamadou2020}. From the perspective of Traffic and Security Engineering, collecting the inbound and outbound traffic of the mobile environment, and further analyzing it, can play a major role to the subsequent optimization and fortification \cite{Conti2018} of both the connected network and the device. Though a plethora of network traffic capturing tools exist for non-mobile devices, with the Wireshark \cite{Orebaugh2007} tool in the lead, mobile devices like tablets and smartphones have access only to a small range of monitoring tools with limited features. 

The process of monitoring network behavior in mobile devices constitutes the need for acquiring information about the received and transmitted information from the given device which can reveal an important knowledge base about the behavior of the network that it is connected to. In this paper, a dedicated mobile network traffic capturing application is presented encapsulating the process of collecting, processing, and analyzing the traffic, while also presenting the user with insightful information such as visual analytics and individual packet characteristics while at the same time offload the computational resources to a remote cloud service that can be deployed on demand by the user. This application was designed with respect to the user's privacy and accessibility and can be used from low to high level of traffic exploration providing a variety of accessible data dependencies such as a pcap traffic format and network flow format. The contribution of this work can be summarized as follows:

\begin{itemize}
    \item \textit{Produces a mobile and easy to use Network Traffic capturing application for Android Devices} 
    \item \textit{Offers a cloud service backbone for traffic processing and archiving} 
    \item \textit{Offers multiple ways of traffic sniffing supporting low to high level firmware privileges}
    \item \textit{Presents a variety of traffic representation and formats}
    \item \textit{Visualizes the captured data in real time for interactive user information} 

\end{itemize}

The rest of this work is organized as follows. Section \ref{Related Work} presents related work performed on the scope of this work. Section \ref{Design and Implementation} describes the architecture, the implementation as well as the features of the CloudCap application. Section \ref{Prototype Deployment} presents the deployment of the application prototype and statistics of its operation. Finally, \ref{Conclusion} summarizes this work.

% ========================================================================================================================
% =====================================================Related Work======================================================
% ========================================================================================================================
\section{Related Work}
\label{Related Work}
Although modern smartphones have significantly increased processing capabilities, they still have limited battery capacity. Computationally heavy tasks can quickly drain the battery of a device, and lead to degradation of user experience. Additionally, smartphone devices often lack many of the features a computer has. For example to capture network traffic the owner of the device has to go through a process known as rooting, resulting often in warranty void. Moreover, Data visualization can undoubtedly help an administrator a lot in finding network anomalies and gaps. Many tools are available that perform this task using the available resources of the smartphone device.
\par
Cloud computing alleviates the device of these computationally intensive tasks \cite{NGUYEN2020133}. Devices can offload such tasks, and be notified  when the processing is completed. This scheme can save a lot of resources, mainly battery though it introduces some additional delay in certain cases. Leung et al. \cite{9373130} developed a cloud based Big-Data analysis platform for COVID-19 pandemic data. Lahmadi et at. \cite{7140443} proposed a cloud based architecture that use the Elasticsearch \cite{ElasticS31:online} NoSQL database and Kibana as the main visualization engine in a cloud server, and an Android agent is responsible for sending log entries to the remote server. Kibana is tasked with presented meaningful visualizations to the user. In \cite{8812144} the authors propose a set of visualization techniques for malware detection. In the hands of the right person these visualizations can lead to finding new threats.
\par
%%% There is a plethora of applications that perform packet capturing and decoding using the device. although this task may be fast
There are also some applications of notice that are available on the Google Play Store \cite{GooglePl55:online} that offer the network traffic capture functionality. For example \textit{Packet Capture} \cite{PacketCa93:online} offers android network traffic acquisition using LocalVPN proxy with ssl decryption and traffic categorization and indexing by app. A disadvantage is that it does not offer any visualization for the sniffed traffic. Furthermore, all the computation and analysis takes place in the mobile device. Another noteworthy app is the \textit{PCAPdroid} \cite{PCAPdroi44:online}, which similarly captures traffic through LocalVPN proxy. This application shows only the captured packets and traffic per application usage and does not offer traffic visualization or computational offloading to the cloud.

%Compared to other readily available applications that similarly to [APPNAME] provide the ability to capture the network traffic from the user’s device, there are some key points that are worth noting. Most of these applications work on non rooted smartphones by utilizing a local VPN to monitor the traffic. They also provide the necessary tools to display the details of the captured packets. However, none of the application reviewed provided a way to preview the overall data  and statistics of the capture in a visual form such as graphs
 %%%%

\section{Design and Implementation}
\label{Design and Implementation}
\subsection{Architecture}
\begin{figure}[h]
    \centering
    \includegraphics[scale=0.5]{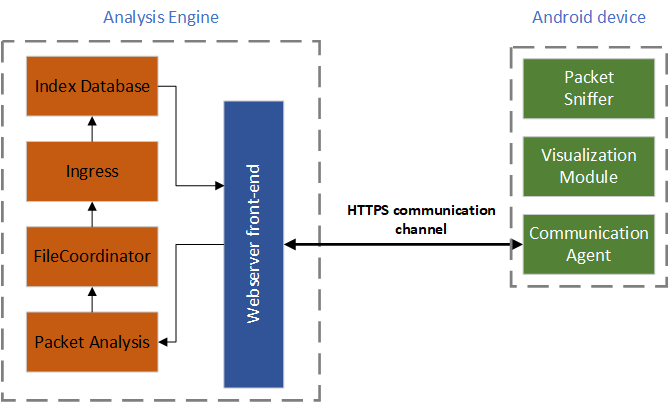}
    \caption{SYSTEM-NAME Architecture}
    \label{fig:systemarchitecture}
\end{figure}

Figure \ref{fig:systemarchitecture} presents a high level overview of the proposed architecture of CloudCap, which consists of two components, the Analysis Engine and the smartphone device. Each component is appointed with tasks that are best suited for it's profile. The Analysis Engine is a cloud application able to perform complex computations and has a wide variety of features. Residing in the cloud, it can scale dynamically and support new functions with little to no effort. CloudCap is a mobile application running in smartphone devices. It is responsible for capturing traffic using filters through a user-friendly interface and send them for analysis to the Analysis Engine. The communication between the two components is based on the HTTPS protocols.
\subsubsection{Analysis Engine}
The Analysis Engine consists of 5 modules. The \textit{Webserver front-end} is the main communication medium between the smartphone application and the Analysis Engine. It provides a REST API for uploading, update on analysis progress and request data. The \textit{Packet Analysis} module is responsible for parsing the captured traffic, extract header fields and store them in an appropriate format the analysis engine can later use. The \textit{File Coordinator} is responsible for coordinating the generated files. This modules moves files in predefined folders and afterward forwards them to the index database. The \textit{Ingress} module is responsible for performing lightweight transformations. Finally, the \textit{Index Database} is responsible for sorting the data and quickly perform look-ups upon receiving a query.
\subsubsection{Smartphone Application}
The smartphone application consists of three modules. The \textit{Packet Sniffer} Module is responsible for capturing traffic. The user, through a Graphical User Interface, specifies the capturing duration and desired protocols. The \textit{Communication Agent} is responsible for communicating with the Analysis Engine. The generated capture files and the server queries and requests are all managed by this module. Finally, the \textit{Visualization module} is responsible for generating the diagrams using the data received by the Analysis engine.

\begin{figure}[h]
    \centering
    \includegraphics[scale=0.55]{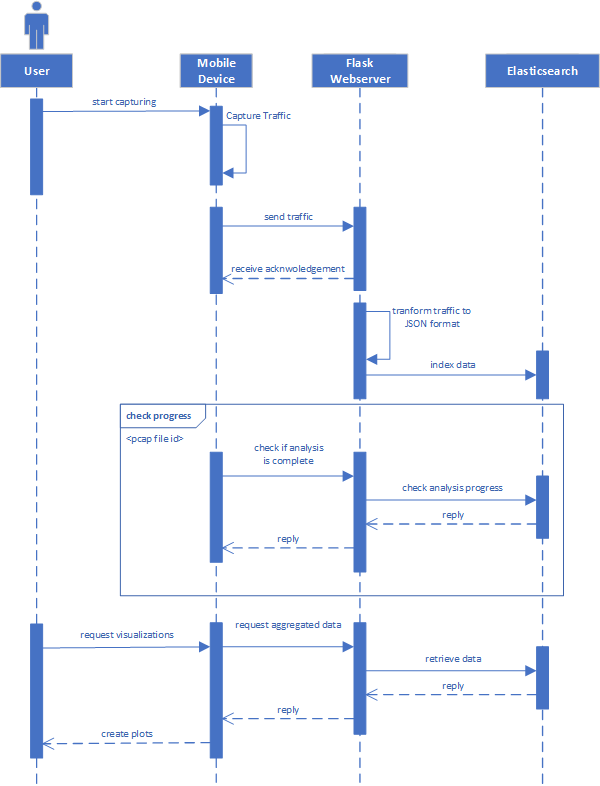}
    \caption{Operation Pipeline}
    \label{fig:communication}
\end{figure}

% ========================================================================================================================
% =====================================================Feature Section====================================================
% ========================================================================================================================
\subsection{Features}
One of the most important parts of a mobile application is its interactivity, versatility and usefulness, which derives from its incorporated features and abilities. Especially for a Network inspection application that aggregates a variety of useful information describing the individual characteristics of the respective network, the functional requirements play a major role in its subsequent practicality. The CloudCap app focuses on visualizing the characteristics of the traffic while allowing for individual packet inspection in different levels of functionality. In particular, the inherent features of the proposed application in this work are as follows.\\

\subsubsection{Network Statistics Visualization}
\begin{figure}[ht]
    \centering
    \includegraphics[scale=0.08]{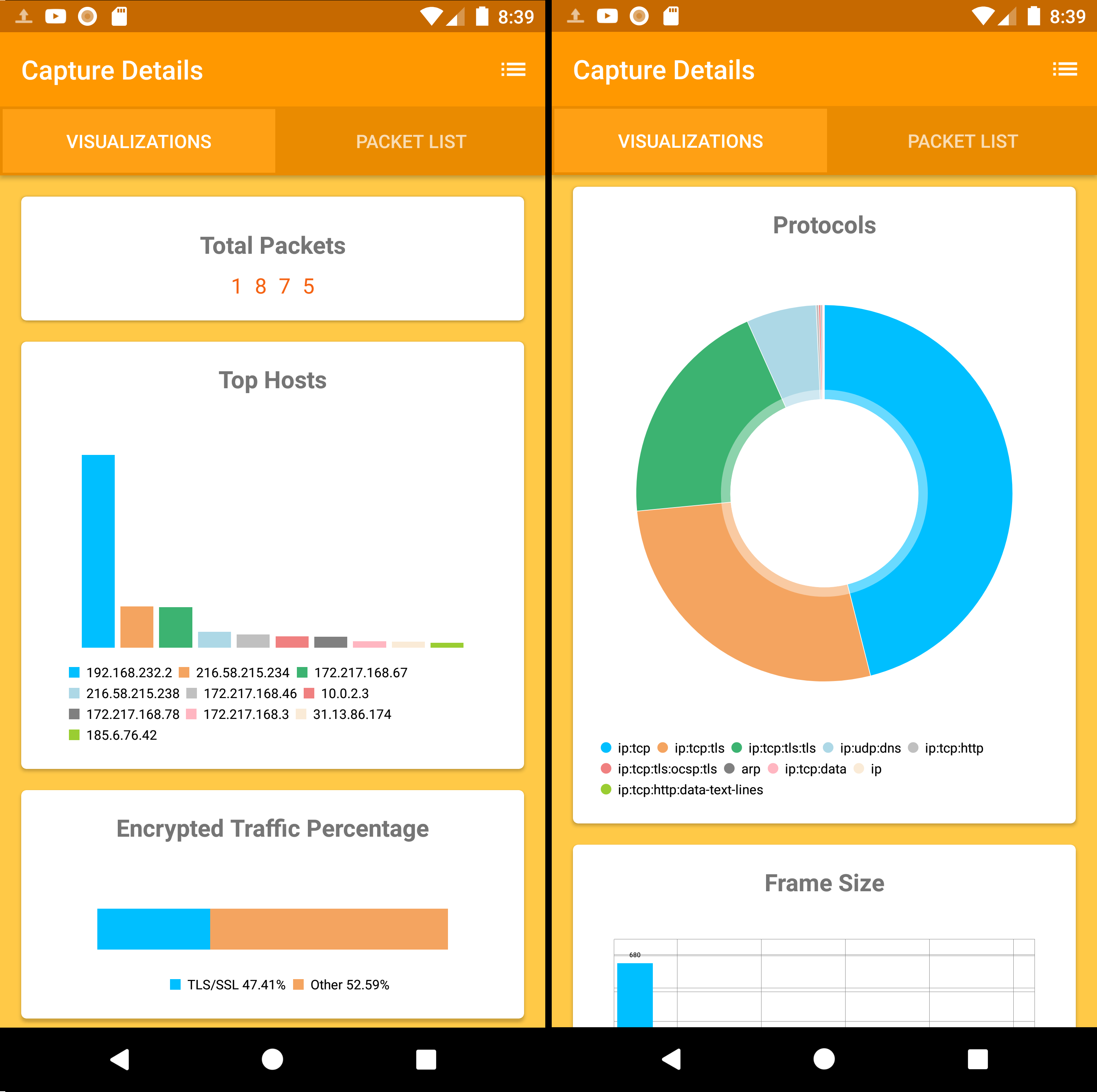}
    \caption{(a) Total Packets, (b) Percentage of Hosts, (c) Percentage of SSL/TLS, (d) Involved Protocols}
    \label{details-4}
\end{figure}

\begin{figure}[ht]
    \centering
    \includegraphics[scale=0.445]{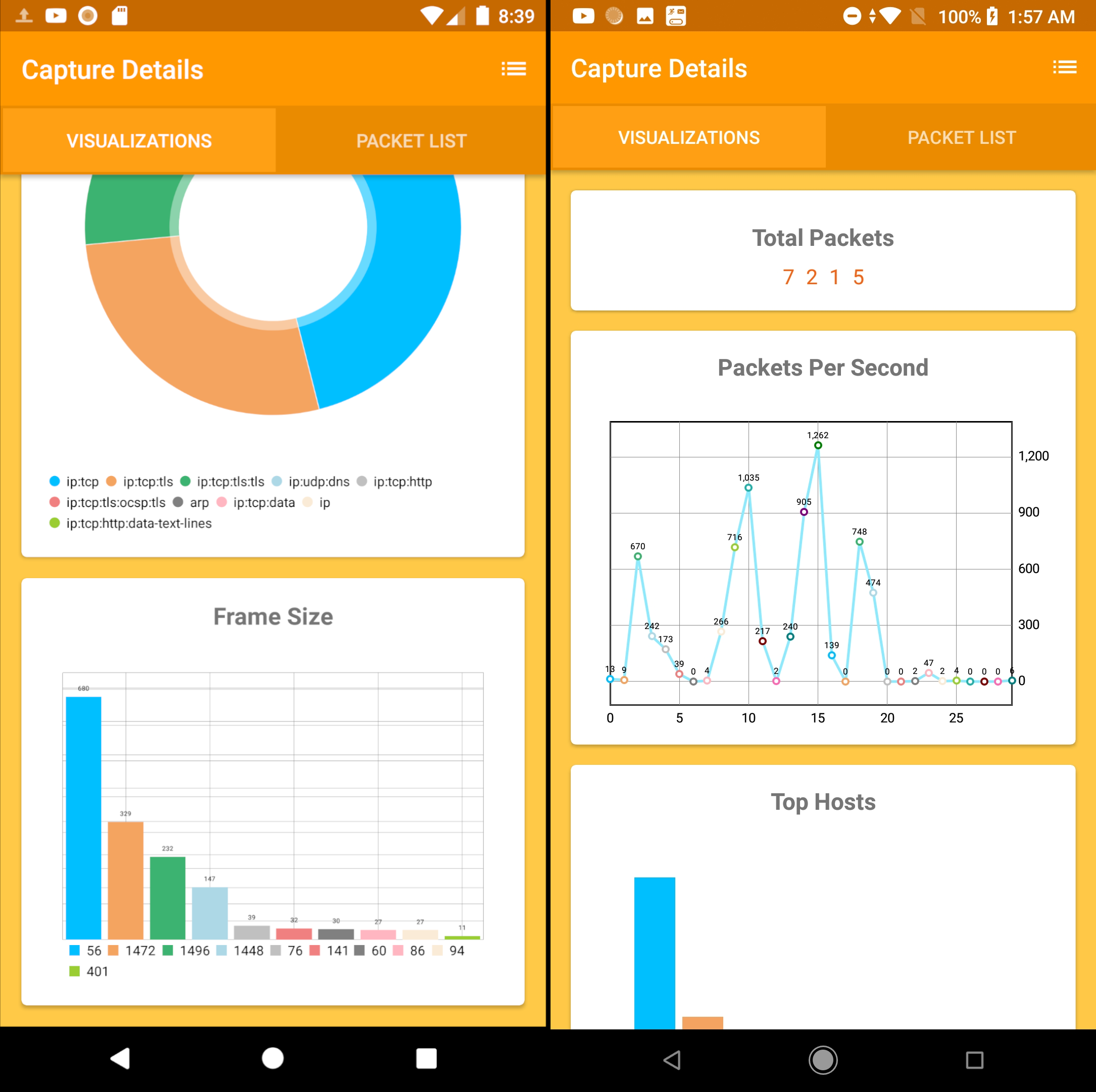}
    \caption{(e) Packet Frame Size, (f)}
    \label{details-3}
\end{figure}

As mentioned, the visualization of the given network traffic plays a major role in the convenience of the application. As so, the presented application offers data depiction through a wide variety of supported visualization. Each visualization presents key elements of the captured traffic, such as, i) Percentage of SSL/TLS packets, ii) Percentage of Hosts, iii) Involved Protocols, iv) Packet Frame Size, v) Packets per Seconds as well as the vi) Total Packets of the captured traffic. Figures \ref{details-4} and \ref{details-3} show examples of the respective visualizations.

\subsubsection{Traffic Archiving}
\label{Traffic Archiving}

\begin{figure}[ht]
    \centering
    \includegraphics[scale=0.7]{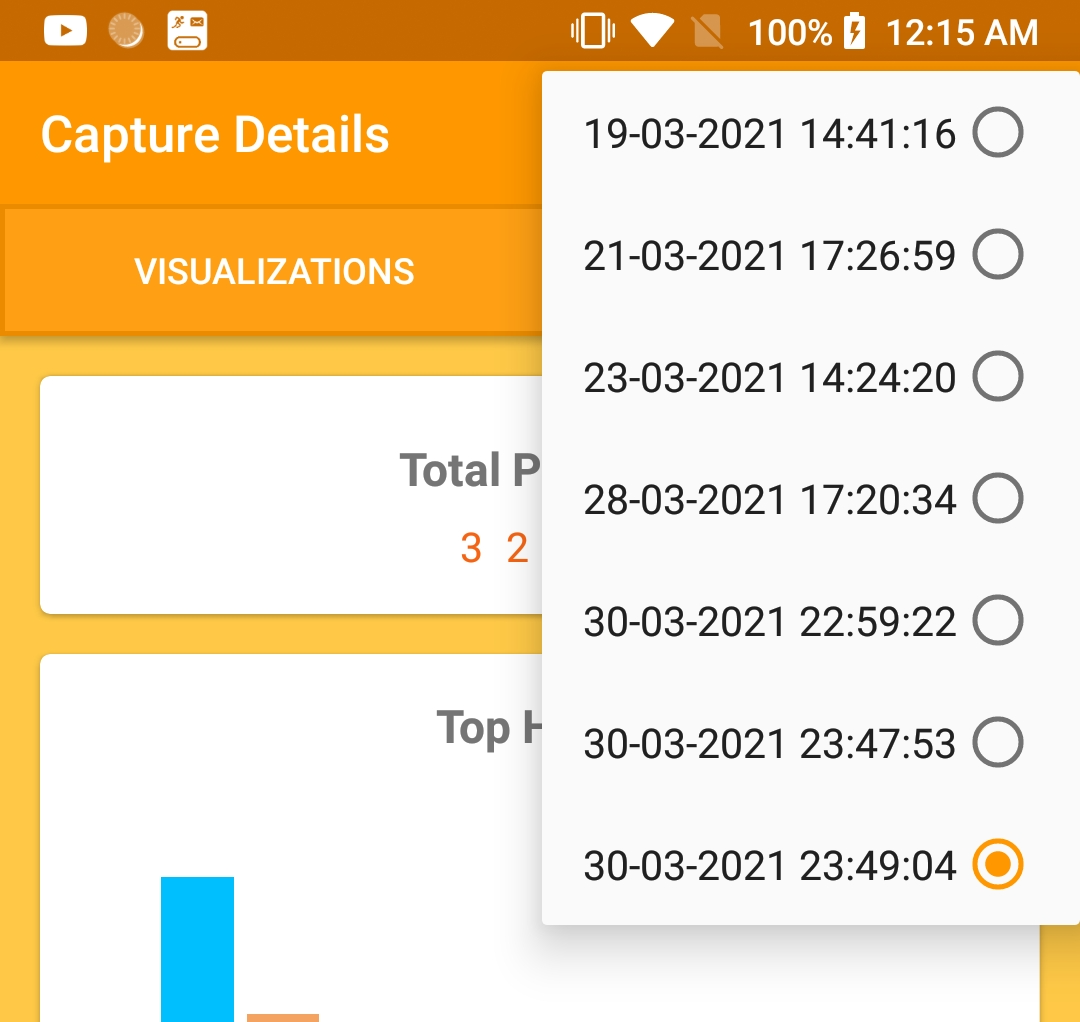}
    \caption{Network Traffic Snapshots}
    \label{details-5}
\end{figure}

As with most of the traffic sniffing applications and platforms, a key function is the archiving of traffic snapshot captured from a specific interface. The generated pcap file, describing the packets collected through the capturing process are first saved locally to the mobile device and subsequently sent to the accompanying server, where they receive an id, analyzed and archived. The archives are called on, on demand from the user, in order to visualize their content, as shown in Figure \ref{details-5}.

\subsubsection{Interactive Packet Inspection}
\begin{figure}[ht]
    \centering
    \includegraphics[scale=0.12]{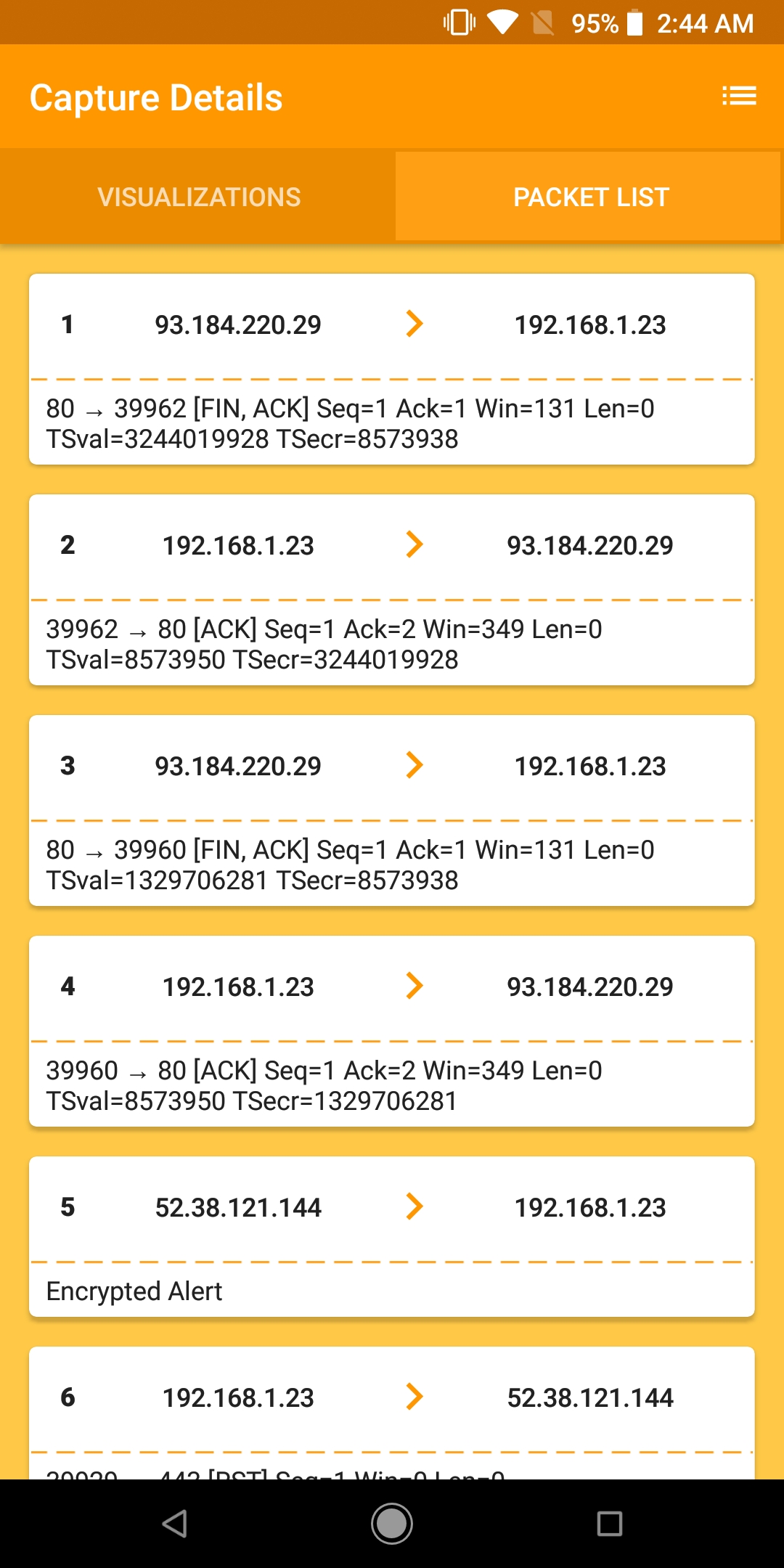}
    \caption{Network Traffic Snapshots}
    \label{details-6}
\end{figure}

Adding to its functionality, CloudCap also incorporates a packet list view for individual packet inspection. As can be seen in Figure \ref{details-6}, the interface depicts the data collected, revealing the source, destination and packet payload information. A view similar to the common Wireshark (internal payload, header and content) was avoided due to the limitation presented by mobile devices in respect to screen size, storage space and touch resolution. 

\subsubsection{Permission Depth Functionality - Root/LocalVPN}
Since most mobile devices do not support promiscuous mode network capability, there are two fundamental ways to capture incoming network traffic. Firstly, the root permission of the phone is utilized. This is conditioned from the availability of the root access of each device, referring to the manufacturers' specifications. In this case, leveraging on-device root terminal level access, executes the pre-compiled tcpdump binary file to successfully sniff the inbound and outbound traffic and redirects it to an output pcap file. This binary file is located the apps data folder so there is no need to be installed in the android system. This method allows filtering using specified flags and interface options so that the traffic capture process matches the users' needs. The second capturing method leverages a LocalVPN loop-back in order to collect the network packets. The drawback of this method is that the packets are read through a byte buffer and the system needs to reconstruct the individual packets. This results to higher computational costs and battery utilization but also reduces the accuracy of the reconstructed packets.

\subsubsection{Network Flow Translation}
To accommodate the need for Traffic Analysis, CloudCap also encapsulates a Network Flow Translation function. A big portion of the traffic analysis performed for network monitoring purposes is done on the network flow level. Especially IDS systems utilizing rule based detection, ML or DL implementations perform intrusion detection and classification on network flows rather than operating based on deep packet inspection, due to the abstraction that the aggregated flows offer. Thus, this application offers the choice of translating the traffic captured on the mobile device to network flows, reducing the complexity of trans-application data transferring and attach a real-time on-demand function. Furthermore, this procedure is overtaken by the server side, thus decoupling the computational cost from the mobile device.

% ========================================================================================================================
% ================================================Prototype Implementation================================================
% ========================================================================================================================
\section{Performance Evaluation}
\label{Prototype Deployment}
In order to evaluate the performance of the presented traffic sniffing application, it was deployed and tested in an evaluation environment. Specifically, the CloudCap application was tested on two actual and one virtual device, a Xiaomi Mi A2, a Nokia 7 Plus and an android emulator respectively. To stress-test the operation of the application the devices captured traffic while streaming HQ videos on the YouTube platform. During the capture operations it was observed that the tcpdump executable utilized $3\%-5\%$ of CPU and $0.1\%$ of memory for a video capture of FHD60fps quality while the uploading process utilized a mean of $11\%$ of CPU and $125Mb$ of memory. While visualizing the results the mean utilization was found to be $0.1\%$ of CPU and $150Mb$ of memory. The resources produced by the application were also measured. In a different test case, with an online radio stream playing in the background for 1 minute, the application captured a total of 2460 packets and generated a 2.5Mb pcap file, while the server analyzed the fetched traffic, producing a 11.2Mb json description (for the visualizations), and supplied the application with the data required to generate the visualizations within 12 seconds. The operation pipeline for the tested system is depicted in Figure \ref{fig:communication} for reference.
% 1. Image of main UI page
% 2. Image Diagrams of an analyzed pcap file.
% 3. 

% ========================================================================================================================
% ========================================================Conclusion======================================================
% ========================================================================================================================
\section{Conclusion}
\label{Conclusion}
As  mobile applications and networking advances, mobile devices are producing more and more data, communicated through mobile networks, such as wireless networks. The need to monitor and process the traffic becomes all the more important in order to effectively optimize and secure both the endpoints and the medium. Unfortunately, there is a limited supply of tools for mobile network traffic acquisition, most of which offer a small number of features. In this work, CloudCap, a new mobile network traffic capturing application is introduced, that undertakes the capturing process while offloading the computational means to a supporting cloud service. CloudCap offers  easy-to-use  and easy to deploy functionality with minimal mobile resource utilization. The proposed app is tested and validated on three experimental environments and present a good amount of utility in mobile wireless networks. 

%Future work of this project constitutes the deployment of the cloud service in a Platform as a Service (PaaS) module, adding Machine Learning and Deep Learning for easy to use intrusion detection, MEC modulation and more.
Future work of this project constitutes the deployment of the Analysis engine in a Mobile Edge Computing (MEC) environment and the addition of Machine Learning or Deep Learning for fast detection of anomalies.

% \section{Acknowledgment}
% This project was developed by members of the UOWM Security Team student group, under University of Western Macedonia.

% \begin{figure}[ht]
%     \centering
%     \includegraphics[scale=1]{figs/logo.png}
%     \label{fig:logo}
% \end{figure}
% 1. Maybe offload them to the edge to reduce communication time
% 2. Automated Analytics through ML

\bibliographystyle{IEEEtran}
\bibliography{references}
\end{document}